# Bipolar doping in van der Waals semiconductor through Flexo-doping

*Bo Zhang[1,2,7], Hui Xia[1,3,7]\*, Zhengdong Huang[1], Yaqian Liu[1], Jun Kang[4]\*, Liaoxin Sun[1,3], Tianxin Li[1,3]\*, Su-Huai Wei[5]\*, Wei Lu[1,3,6]\**

[1]State Key Laboratory of Infrared Physics, Shanghai Institute of Technical Physics, Chinese Academy of Sciences, Shanghai 200083, China.
[2]Department of Physics, Shanghai Normal University, Shanghai 200234, China.
[3]University of Chinese Academy of Sciences, Beijing 100049, China.
[4]Beijing Computational Science Research Center, Beijing 100193, China.
[5]Eastern Institute of Technology, Ningbo 315200, China.
[6]School of Physical Science and Technology, ShanghaiTech University, Shanghai 201210, China.
[7]These authors contributed equally: Bo Zhang, Hui Xia.

\*Corresponding author: huix@mail.sitp.ac.cn (H. Xia); jkang@csrc.ac.cn (J. Kang); txli@mail.sitp.ac.cn (T. Li); suhuaiwei@eitech.edu.cn (S. Wei); luwei@mail.sitp.ac.cn (W. Lu).

**Abstract**

Doping plays a key role in functionalizing semiconductor devices, yet traditional chemical approaches relying on foreign-atom incorporation suffer from doping-asymmetry, pronounced lattice disorder and constrained spatial resolution. Here, we demonstrate a physical doping technique to directly write nanoscale doping patterns into layered semiconductors ($MoS_2$). By applying localized tensile and compressive stress via an atomic force microscopy probe, p and n type conductance are simultaneously written into the designed area with sub-100-nm resolution, as verified by spatially resolved capacitance and photocurrent experiments. Density functional theory calculations reveal strain-driven shifts of donor and acceptor levels, as large as several hundreds of meV, linking mechanical stress to semiconductor doping. Fabricated strain-engineered junction efficiently rectifies the current flow and performs logic operations with stable dynamic response. This strain-driven approach enables spatially precise doping in van der Waals materials without degrading crystallinity, offering a versatile platform for nanoscale semiconductor devices.

Doping is a foundational pillar of semiconductor manufacturing. Conventional doping techniques, which employ chemical processes to incorporate foreign atoms into semiconductor lattices, face significant challenges such as asymmetric doping behavior[1-6], pronounced lattice disorder[7,8], and constrained spatial resolution[8]. For instance, in layered semiconductors like $MoS_2$, achieving n-type conductivity is relatively straightforward[9-12], yet developing reliable p-type doping strategies for this two-dimensional material remains a major hurdle[13-20]. Current research indicates that, aside from degenerate substitutional doping with Nb atoms[13-17], most existing methods struggle to overcome the material's intrinsic doping tendencies and enable effective p-type conversion[21-25]. Moreover, traditional chemical doping unavoidably introduces structural defects that severely degrade critical electronic properties[8]. Additionally, dopant diffusion during chemical doping often causes lateral spreading of the doped regions, sometimes extending up to several micrometers[26-28].

Strain engineering could be an efficient way to modulate the doping of semiconductors. As presented in literatures, the strain could increase the solubility of the donor As and Sb in Si by several times[29-31]. It is interpreted as the offset of the internal compressive/tensile stress (introduced by the volume difference between the dopant and the host atom) by the external tensile/compressive stress, which thus modifies the formation and activation energies of the donor[32-35]. Based on this principle, there is a chance to make use of the strain process to selectively activate donor or acceptor and thus turn a semiconductor into n or p type conductance. However, to date, no such route has been demonstrated.

In this work, we show a strain-mediated flexo-doping technique to directly write nanoscale doping patterns into layered semiconductor. Specifically, an atomic-force-microscopy (AFM) was used to apply tensile or compressive stress to the nanoscale area of a layered semiconductor. By this effort, the defect formation energy and activation energy of that specific region can be tuned in a large extend and in a flexible way. Taking the layered $MoS_2$ as an example, the defect transition energy levels are shifted toward the valence band maximum under tensile stress, thus the donors are deactivated and acceptors are activated. It finally turns the $MoS_2$ layer from n-type to p-type doped, with sub-100-nm in minimum linewidth. The findings disclosed here might provide a feasible path to design nanoscale doping pattern in the layered semiconductor for diverse functionality.

Strain could selectively enhance the doping characteristic of semiconductor. As theoretically revealed by Zhu et al. [34] and more recently by Yan et al.[32]: if the dopant has large size than the replaced host element, when a tensile strain is applied, the formation energy of the dopant created defect is reduced. The reduction is, in general, proportional to the size differences and the applied strain. Considering that the negatively charged defect usually is larger and the positively charged defect is usually smaller than the corresponding neutral one, when a tensile strain is applied, the transition energy level is pushed down in energy, and in most cases reduces the transition energy level with respect to the VBM. That is, when tensile strain is applied to the system, the system can be more p-type; and when a compressive strain is applied, it will be more n-type.

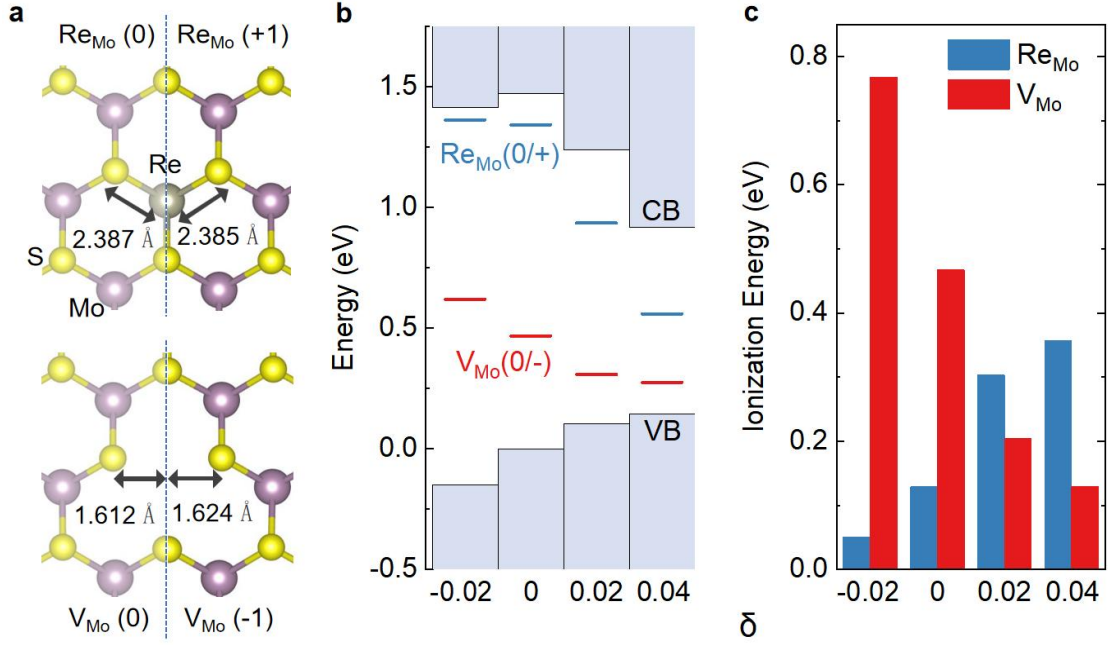

**Fig. 1| Strain-dependent doping behavior in MoS$_2$. a,** Atomic structure of Re$_{Mo}$ donor and V$_{Mo}$ acceptor before and after ionization. **b,** The calculated band alignment, Re$_{Mo}$ (0/+) donor level and the V$_{Mo}$ (0/-) acceptor level under different in-plane strain δ. **c,** The ionization energies of Re$_{Mo}$ and V$_{Mo}$ under different δ.

Previous studies have shown that mechanical strain can enhance dopant solubility in semiconductors, thereby increasing charge carrier concentration. However, evidence for strain-induced reversal of semiconductor doping polarity remains elusive. To investigate this possibility, we conducted density functional theory (DFT) calculations to systematically evaluate strain-mediated effects on donor and acceptor ionization energies, using MoS$_2$ as a representative model system. For the donor, here we considered Re substitution of Mo (Re$_{Mo}$), which is considered as a possible origin of the n-type conductance in MoS$_2$[36]. For the acceptor, we considered Mo vacancy (V$_{Mo}$), which is an intrinsic acceptor[36-38]. Details of the calculations are given in the Methods section. Fig. 1a shows the atomic structure of Re$_{Mo}$ donor and V$_{Mo}$ acceptor before and after ionization. Obviously, the ionization of Re$_{Mo}$ donor and $V_{Mo}$ acceptor would increase and decrease their size, respectively, in consistent with the previous expectations[32,34]. In this regard, we could apply an external stress to either counteract or enhance the internal one and thus suppress or promote the defect ionization. The DFT calculations support such judgement. In Fig. 1b, the (0/+) donor levels and the (0/-) acceptor levels under different in-plane strain δ of $MoS_2$ are plotted. When δ is 0, the donor level is shallow and the acceptor level is deep, consistent with the n-type conductance. As δ increases (means that $MoS_2$ is increasingly stressed by external tensile force), both the donor and acceptor levels shift downwards. As a result, the ionization energy of the donor increases with increasing δ, while that of the acceptor decreases (Fig. 1c). This is in good agreement with the above discussions. When δ reaches 0.04, the donor level is close to the gap center, whereas the acceptor becomes quite shallow, with an ionization energy of 0.13 eV. Hence, the $MoS_2$ could transform from n-type to p-type. It should be emphasized that i) the tensile stress induced bandgap narrowing would accelerate this transition process since the valence band is lifted that helps further reduce the ionization energy of acceptor; ii) here we are not claiming that $V_{Mo}$ is the sole intrinsic defect to introduce p-type behavior in $MoS_2$. The purpose of the DFT calculations is to show the trends of the donor and acceptor levels in strained $MoS_2$, and this trend is universal for different point defects in different semiconductors[32-35].

Following the theoretical validation of strain-mediated doping, the challenge of devising a practical experimental implementation arises. To address this, we introduce an atomic force microscopy (AFM) probe-assisted strain-doping technique, which we term the flexo-doping approach. Fig. 2a schematically shows the device structure and its preparation procedure. Typical lithography and etching processes were performed to fabricate patterned holes on a 280 nm-SiO$_2$/350 μm-Si substrate. The hole can be carved into any shape, and herein we show the case of equilateral triangle, with ~15 nm in depth and ~4 μm in the side length. After that, a uniform layered $MoS_2$ film was transferred onto the structured substrate by a standard dry-transfer procedure. In such device, the flexible nature of $MoS_2$ film makes itself attach to the substrate and thus exhibit a concave morphology. This is critical to the subsequent physical doping process. Fig. 2b schematically shows the process of writing and measuring the nanoscale doping pattern on such device. At the writing stage, a diamond-coated silicon probe is pressed hard onto the $MoS_2$ surface. The contact press force is up to +728 nN, as indicated in the atomic-force-microscope force-distance curve. When the probe scans from right to left, this force allows the probe to drag and squeeze the $MoS_2$ lattice hardly, and then give rise to tensile and compressive deformation at the right and left sidewalls, respectively. At the measuring stage, the probe is controlled to nearly detach from the sample surface (adhesion force region, -70 nN), thus would not destroy the stress status of the sample. This is verified in variable angle measurements, as shown in Supplementary Fig. 1. Meanwhile, the probe is set virtually grounded and a 1V 90 kHz AC bias is supplied to the electrode deposited on the $MoS_2$ surface. By this effort, the probe can serve as a mobile electrode to measure the differential capacitance (dC/dV) point by point, and therefore show the doping distribution of the device with nanoscale resolution. This measure is known as scanning capacitance microscope mode (SCM) that has been well demonstrated in the silicon and III-V semiconductor industry. More details could be found in Supplementary Fig. 2.

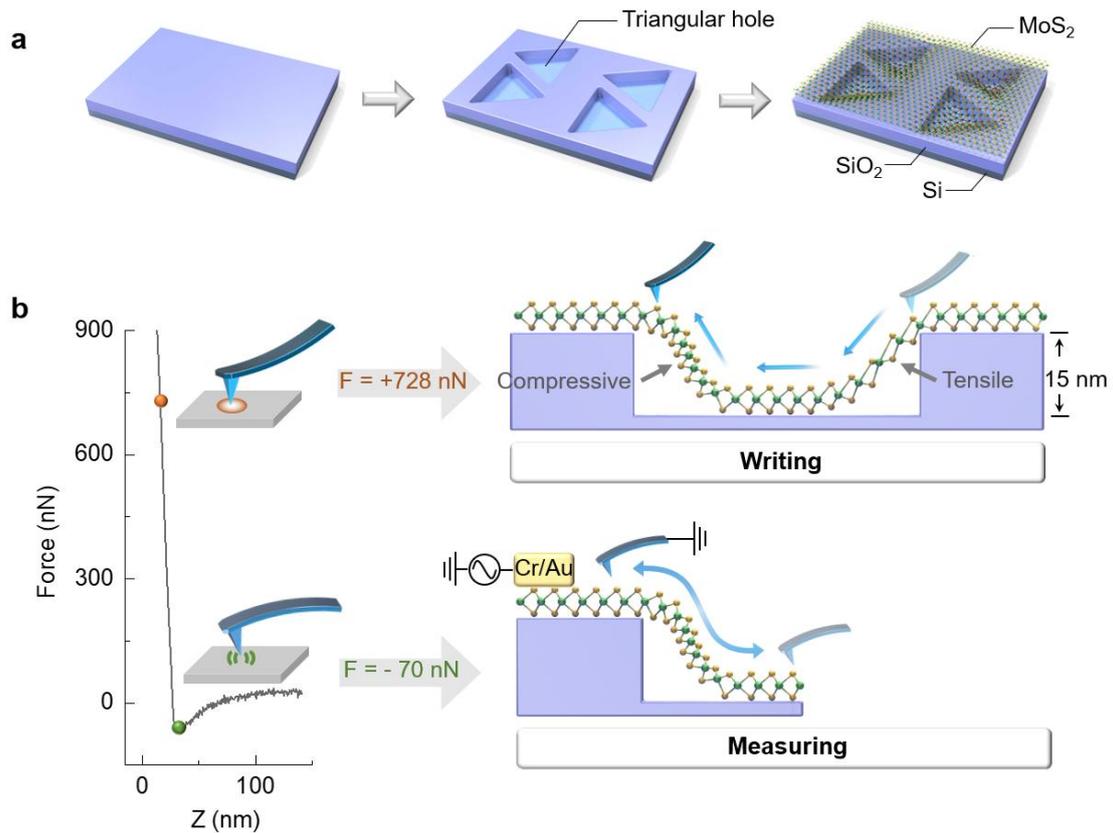

**Fig. 2| Strategy of flexo-doping of MoS$_2$. a,** Device structure and the preparation procedure. Lithography and dry-etching processes were used to etch hole structures in the SiO$_2$/Si substrate. After that, a uniform MoS$_2$ film was transferred onto the substrate. **b,** Process of writing and measuring nanoscale doping pattern on the MoS$_2$ device. At the writing stage, the probe works in the repulsive force region, +728 nN, as marked with the brown ball in the atomic-force-microscope force-distance curve. The probe is then capable of dragging and squeezing the MoS$_2$ crystal at the two sidewalls, respectively, as it scans over the sample surface. This will write tensile and compressive stress and finally change the doping polarity and concentration of these two locations. At the measuring stage, the probe works in the adhesion force region, -70 nN, as marked with the green ball in the atomic-force-microscope force-distance curve. It makes sure that the measuring process would not destroy the strain status of the sample. During the measurements, the probe detects the differential capacitance signals of the MoS$_2$ film point by point, which finally depicts the doping pattern of the device in nanoscale resolution. The probe is kept virtually grounded, and an AC bias is applied to the Cr/Au electrode.

The writing direction determines which side to apply tensile or compressive stress. As depicted in Fig. 3a, when the probe scans along the median of an equilateral triangle hole from right to left, the tensile stress is written into the right two sidewalls and the compressive stress is written into the left single sidewall. By contrast, when the probe scans along the bottom side of a triangle hole, the bottom, right and left side will be in stress-free, tensilely stressed, and compressively stressed status, respectively. These two strain writing modes would lead to completely different doping patterns. As shown in Fig. 3b, 7 L and 9 L $MoS_2$ are transferred onto two triangle holes, respectively. And the strain writing process just follows the schemes shown in Fig. 3a. Generally, (i) the flat area shows a uniform negative dC/dV signal in both devices. It is the response of majority electrons to the AC bias voltage and thus confirms the intrinsic n-doped behavior of few layer $MoS_2$[39]. (ii) The first scheme dopes the right two sidewalls of the triangle hole into the p-type conductance, that is characterized by a positive dC/dV response, ~11 mV. Also, it dopes the left single side into $n^+$-type conductance, that is featured by a larger negative dC/dV response as compared with that of the flat region, ~-77 mV *vs* ~-40 mV (Fig. 3c). (iii) In the second scheme, the bottom sidewall is kept in n-type conductance, the right sidewall is turned to p-doped, and the left sidewall is turned to $n^+$-doped. It is worth noting that bubbles, labeled as I, II, III, IV, V in Fig. 3b, are commonly found in the devices due to the air residue during the dry-transfer process. And the flexo-doping effect could also be observed in these structures. Such behavior is discussed in more details in the supplementary material.

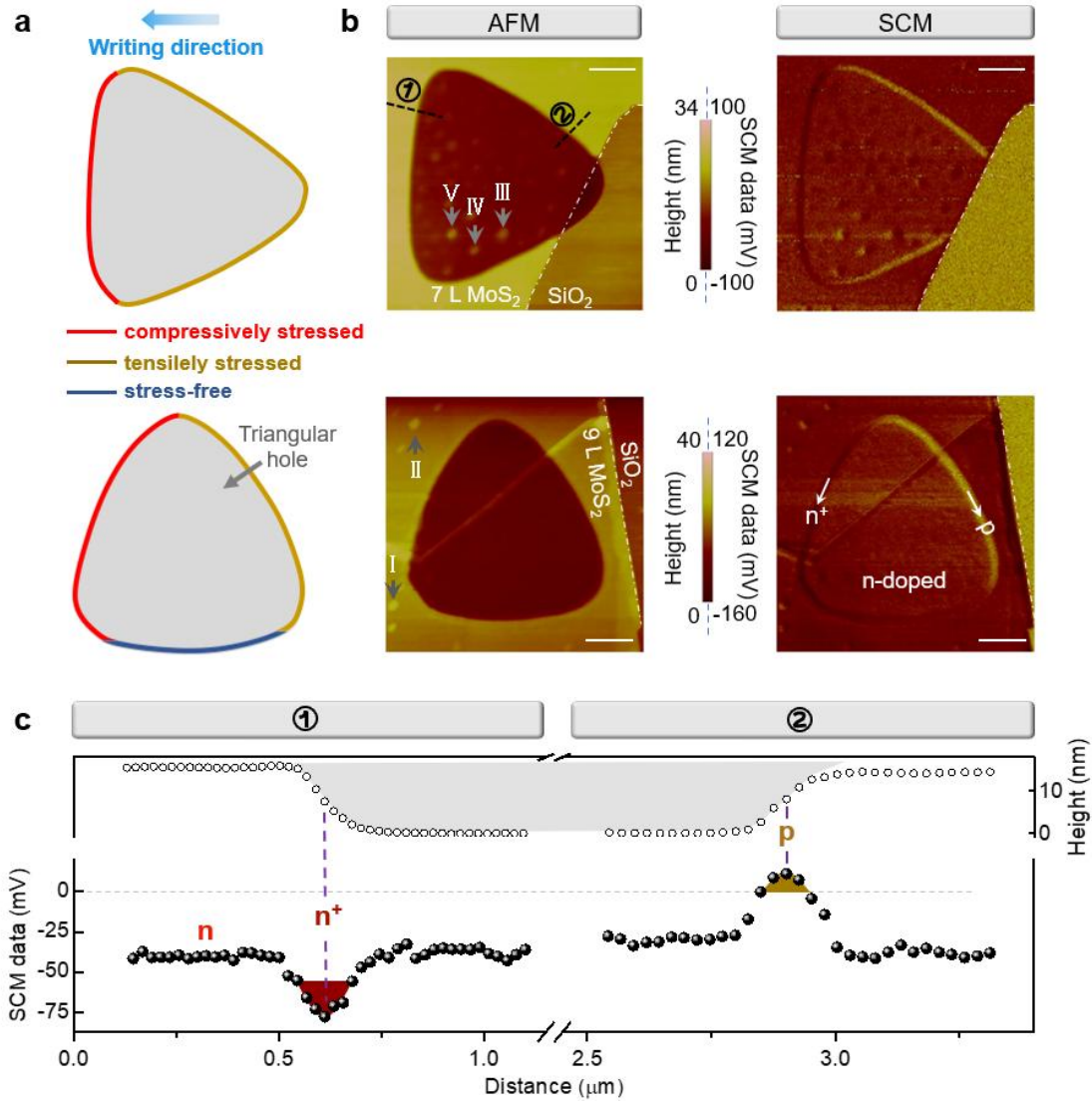

**Fig. 3| Direct writing nanoscale doping-pattern into MoS₂. a,** Schematically showing the writing direction and its influence on the stress distribution. In the first scheme, the probe scans along the median of an equilateral triangle hole from right to left. The tensile stress is then written into the right two sidewalls and the compressive stress is written into the left single sidewall. In the second scheme, the probe scans along the bottom side of an equilateral triangle hole. The bottom, right and left side will be in stress-free, tensilely stressed, and compressively stressed status, respectively. **b,** The output doping-pattern. The scanning capacitance microscope images show the doping distribution of the devices in nanoscale resolution, while the atomic force microscope images give the corresponding morphology information. In the device doped in the first scheme, the MoS₂ film is 7 L thick, and the doped area is typically ~100 nm in width. In the device doped in the second scheme, the MoS₂ film is 9 L thick, and the doped area is typically ~90 nm in width. It is worth noting that bubbles are commonly found due to the air residual and would be written into stress too. Some bubbles are marked out with gray arrows and numbered I-V. The white dash-dot lines outline the boundary between MoS₂ and SiO₂. The scale bars are 1 μm. **c,** Height and differential-capacitance profiles of the nanoscale-

doped device. The data are extracted along the two cutting lines marked as ① and ② in **b**. The flat MoS$_2$ is intrinsically n doped, that is featured by a negative differential capacitance, ~-40 mV. By contrast, the left sidewall is n$^+$ doped, evidenced by a very negative differential capacitance signal, ~-77 mV; the right sidewall is p doped, evidenced by a positive differential capacitance signal, ~ 11mV.

To test the flexo-doping effect further, we prepared a unipolar strained device and performed the scanning photocurrent microscope experiment (SPCM) on it. As shown in Fig. 4a and b, a MoS$_2$ flake was transferred onto a stepped SiO$_2$/Si substrate, with half of the material upstairs and the other half downstairs. After that, symmetrical Au electrodes were deposited on the two stairs, respectively. In such device, we didn't apply additional strain to the device via probe scanning but keep the status as it originally was. Considering that the vertical drop is up to 100 nm, the intrinsic tensile stress is sufficiently high to lower the background electron concentration at the sidewall and even turn it to p-type doped. This is proved by the SCM experiments, as shown in the Supplementary Fig. 3. Fig. 4b shows the SPCM image of the unipolar strained device (see experimental details in Methods section). Obviously, there are four photocurrent hot lines in the device, marked as A, B, C and D, respectively. Features A and D arise from an efficient separation of photo-generated carriers at the large Schottky barrier located at the MoS$_2$/Au interface[40]. The different signs of photocurrent indicates that the photocarriers generated in these two locations would flow in an opposite direction. This is reasonable since the two Schottky junctions naturally have their back toward each other. Features B and C are also back-to-back and centered at the morphology transition line. It indicates that the vertical drop and the resulting tensile stress benefit the photocarrier collection significantly.

Considering that the magnitude approaches that of Schottky junction, the back-to-back photoresponse observed here should be associated with the strain doping junction.

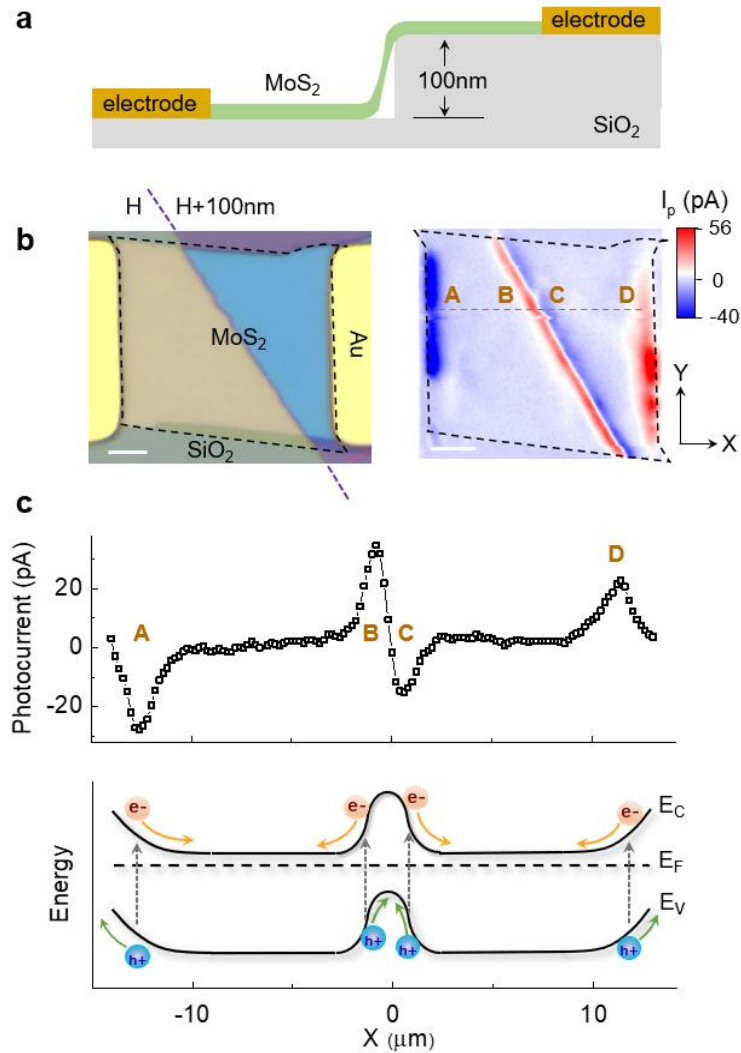

**Fig. 4| Charge carrier separation at the flexo-doped junction. a,** Schematically showing the device structure. SiO$_2$/Si substrate was first etched into a stepwise morphology. The height difference between the two stairs is ~100 nm. A MoS$_2$ flake was then transferred onto the substrate, with half of the material upstairs and the other half downstairs. After that, symmetrical Au electrodes were deposited on the two stairs, respectively. **b,** Optical microscope and scanning photocurrent microscope images of the device. Four distinct features are labelled as A, B, C and D in the scanning photocurrent microscope image. The black dash lines outline the boundary of MoS$_2$. The purple dash line indicates the location and heading of the boundary between the two stairs. "H" and "H+100nm" represent the height of the two stairs. The scale bar is 5 μm. **c,** Photocurrent profile and band alignment. The photocurrent profile was extracted along the gray dash line as shown

in **b**. There are two Schottky junctions and a strain doped npn junction. The two Schottky junctions have their back toward each other, thus collect photocarriers in opposite directions. This behavior leads to a positive and negative photocurrent signal at A and D locations, respectively. The npn junction is formed by the tensile strain. It is featured by the back-to-back photocarriers harvest, which results in a negative and positive photocurrent signal at B and C locations, respectively.

Fig. 4c gives an intuitional image of such distinct behavior. Typically, the band bending of Schottky junctions drives the photo-excited electrons into the device channel and sweeps the photo-excited holes toward the contact. This law leads to the negative and positive photoresponse at the A and D locations, respectively. By contrast, the intrinsic tensile stress turns $MoS_2$ into $n^-$ or p type conductance at the morphology transition line. This naturally forms a $nn^-n$ or npn junction in the conduction channel and thus contributes to the back-to-back photoresponse. Finally, the Schottky and strain induced junctions lead to three times switching of the sign of photocurrent between the four distinct features (A, B, C and D).

As stated above, the strain easily introduces p-type edge-doping to $MoS_2$ film (more doping patterns written in different shapes and directions are shown in Supplementary Fig. 4-7), the linewidth of which is proved down to ~90 nm. This is a significant achievement since p-type doping in $MoS_2$ is very difficult using other approaches. Next, we would show the case of flexo-area-doping, which could expand the linewidth of doping area to several hundred nanometers. As schematically shown in Fig.5 a, patterned holes with ~300 nm in width and ~100 nm in depth were fabricated on $Si/SiO_2$ substrate. After that, a $MoS_2$ film was transferred onto the prepared substrate by standard dry-transfer process. Considering that the submicron holes are in high depth-to-width ratio, part of MoS$_2$ film is therefore suspended. Such characteristic offers an opportunity to conveniently introduce tensile stress and then p type doping to the whole area, e. g. by utilizing AFM probe to squeeze the suspended film into the hole. SCM measurements help record the whole process. As shown in Fig.5 b, the probe force is low at first, ~15 nN, the MoS$_2$ film then exhibits a uniform n type doping behavior. And with the increase of probe force to 78 and 312 nN, an obvious doping reversal was observed at the edges of triangular holes. Finally, when we increase the probe force to 390 and 469 nN, the whole suspended film was turned into p type doping.

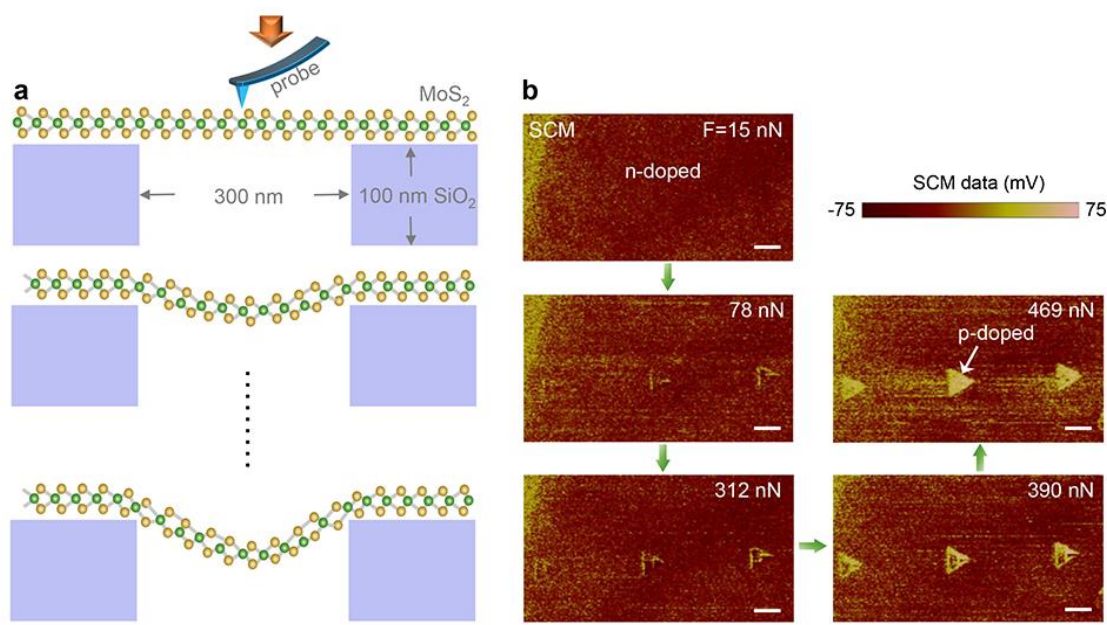

**Fig. 5| Area-doping of MoS$_2$ film. a,** Schematically showing the process to realize area-doping of MoS$_2$ film. Part of MoS$_2$ film is suspended on holes with high depth-to-width ratio. AFM probe was then used to squeeze the suspended MoS$_2$ film into the holes. By this means, tensile stress and thus p type doping are introduced to the whole suspended MoS$_2$ film. **b,** SCM images of the MoS$_2$ film under different probe forces. Initially, the whole film is suspended on the holes and exhibits n type conductance. And with the increasing of probe force, the suspended MoS$_2$ film was squeezed into the holes. This leads to the doping reversal, first on the edges of the pattern and then extending to the whole region. The scale bar is 500 nm.

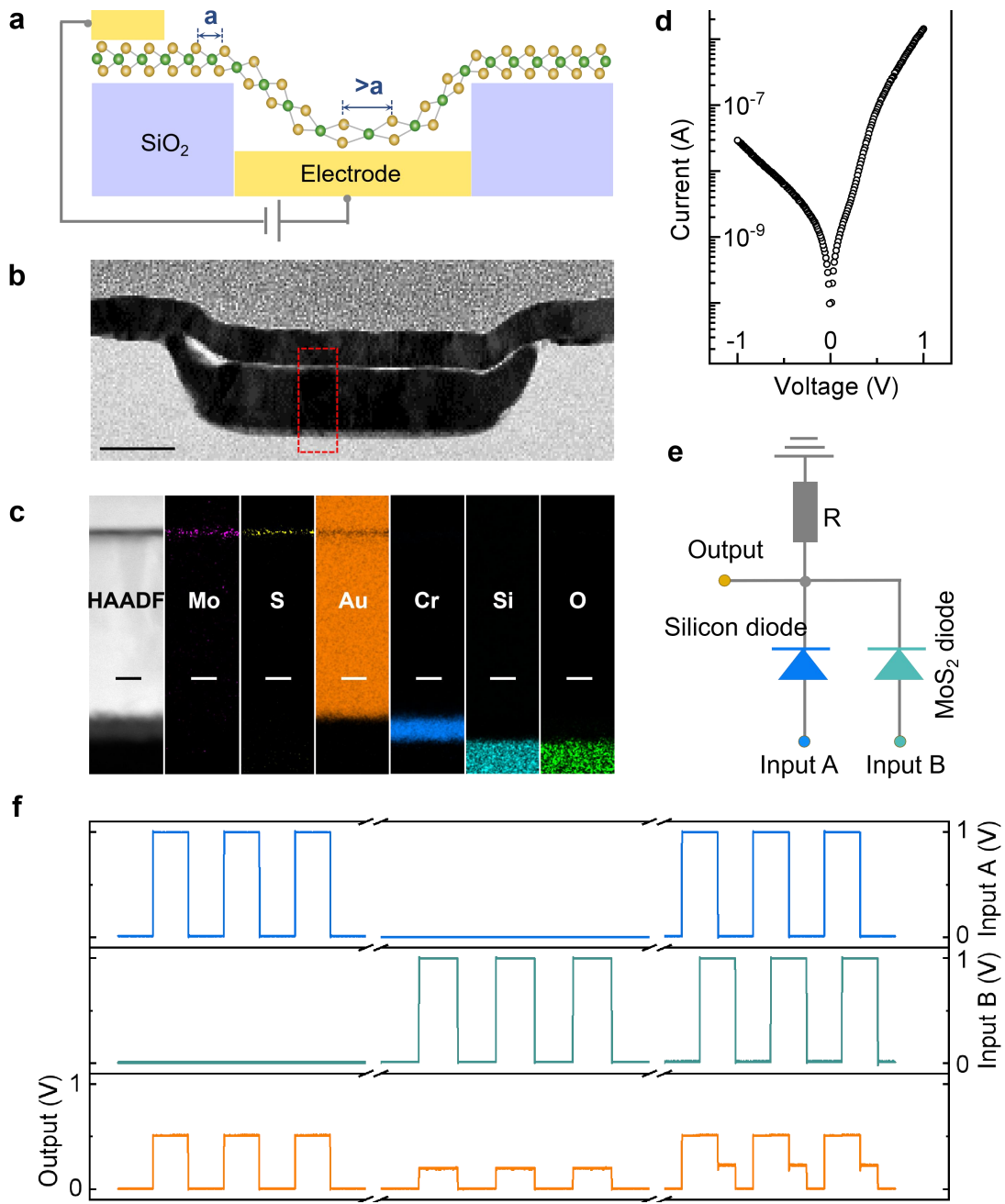

**Fig. 6| Flexo-doped MoS$_2$ junction and its current and voltage output. a,** Schematically showing the structure of MoS$_2$ junction fabricated by flexo-doping. **b,** Bright-field transmission electron microscopy (BF-TEM) image of the cross section of the flexo-doped MoS$_2$ device. The scale bar is 50 nm. **c,** High-angle annular dark-field transmission electron microscope (HAADF-TEM) and energy-dispersive X-ray spectroscopy (EDS) images of the flexo-doped region as marked by the red box in **b**. Note that an additional Au film (~20 nm) was deposited on the MoS$_2$ film for the TEM experiments. It is expected to minimize the lattice damage from the focused-ion-beam etching process. The scale bar is 5 nm. **d,** IV curve of the flexo-doped MoS$_2$ junction. **e,** Logic OR gate based on a flexo-doped MoS$_2$ junction and a commercial silicon diode. **f,** Dynamic output voltage response of the OR gate.

The flexo-area-doping strategy allows us to conveniently fabricate 2D pn junction and make use of it in electronic applications. As schematically shown in Fig. 6a, a MoS$_2$ film transferred onto a sub-micron hole naturally forms a pn junction. Specifically, the MoS$_2$ segment fallen into the hole is tensilely stressed and thus turned into p-doped, while the other part is stress-free and n-doped. To make sure the electrical contact of the fallen MoS$_2$ segment, an Au film was deposited into the hole ahead (Fig. 6b and c). More details of the device fabrication process could be found in the Methods section and Supplementary Fig. 8. Fig. 6d shows the IV curve of the flexo-doped MoS$_2$ junction, in which an obvious current-rectifying behavior is observed with the rectification ratio close to two orders of magnitude. To further evaluate its rectifying performance, we integrate the flexo-doped MoS$_2$ junction into a logic device. Fig. 6e shows the structure of a OR gate, which consists of a commercial silicon diode (IV curve is shown in Supplementary Fig. 9), a flexo-doped MoS$_2$ junction and a 1 MΩ resistor. The dynamic output voltage response of the OR gate was investigated by supplying 1 V square pulse signals to the two input ends. Obviously, the flexo-doped MoS$_2$ junction works well since it steadily outputs a square signal, just like that done by the silicon diode (Fig. 6f). It allows the OR gate to successfully perform logic operations. Note that the high voltage level is ~0.2 V and ~0.5 V for MoS$_2$ junction and the silicon diode, respectively. This difference arises from the impedance mismatch between the two kinds of diodes.

- **Conclusion**

In summary, we propose and demonstrate an approach to overcome doping difficulty and directly write nanoscale doping pattern into a van der Waals material. In this approach, a nanoscale probe is used to apply tensile or compressive stress to the designed areas of a van der Waals 2D material such as $MoS_2$, which will subsequently give rise to hole or electron conductance at that local area (sub-100-nm in minimum linewidth). Such distinct behavior is explained by the flexo-doping effect, in which intrinsic donor or acceptor like defects are selectively activated by shifting their energy levels toward the band edge of $MoS_2$ via a controlled strain process.

The proposed approach to tune the defect activation energy by local strain is universal[4], which is independent on the crystal structure, component, and interlayer bonding force of semiconductors[33-35]. Thus, the doping strategy disclosed here is applicable to other semiconductors when they are thinned to be flexable, especially when it is monolayers' thick. Also, considering that this approach uses only existing defects already in the system, it does introduce more scattering centers. Those characters indicate that the flexo-doping approach might greatly simplify the fabrication process and enhance the performance of nanoscale semiconductor devices.

**Acknowledgments**

This work was supported by the Strategic Priority Research Program of the Chinese Academy of Sciences (Grant Nos. XDB0580000), National Natural Science Foundation of China (Grant Nos. 12174416; 12393833; U2241219; 12227901, U23A6002, 12393831 and 12088101), the National Key Research and Development Program of China (Nos. 2022YFA1404603, 2024YFA1409802).

**Author Contributions**

H. Xia, J. Kang, T. Li, S.-H. Wei and W. Lu supervised the project. H. Xia proposed the idea and designed the experiments. B. Zhang, Y. Liu and Z. Huang fabricated the device. H. Xia, B. Zhang and L. Sun performed the (opto-) electrical experiments. J. Kang carried out the DFT calculations. H. Xia analyzed the data and prepared the original manuscript. Everyone involved in discussion and modifying the manuscript.

**Conflict of interest**

The authors declare no competing interests.

**Data Availability**

All the data that supports the plots within this paper and other findings of this study are available from the corresponding author upon request.

**METHODS**

**Sample preparation for stress writing.** The sample for flexo-doping was fabricated by standard lithography, etching, and dry-transfer processes on a Si/280-nm-SiO$_2$ substrate. First, triangle pattern, with ~4 μm in the side length, was depicted by a lithography process performed in a MicroWriter ML3 machine (Durham Magneto Optics). Second, the triangle pattern was transferred onto the substrate by a reactive ion etching process in a PlasmaProSystem 100 system. A mixture of CHF$_3$ (40 sccm) and O$_2$ (6 sccm) was selected as the etching gas and the etching power was set as low as 100 W. These configurations help reduce the etching damage and surface roughness. The etching depth is designed and controlled to be ~15 nm. Third, a uniform MoS$_2$ film was transferred onto the patterned substrate by a standard dry-transfer process.

**The stress writing direction.** In AFM based experiments, the probe scans twice at each fast-axis line, first left-to-right and then right-to-left. The two scanning trajectories are known as trace and retrace processes. Generally, the trace scan is soft and the retrace scan is hard, which leads to an overall right-to-left lateral force. Such characteristic is verified in an AFM assisted surface cleaning experiment. As shown in Supplementary Fig. 10, the probe was kept scanning over a MoS$_2$ sample that is full of photoresist residual covered. The probe-sample press force was tuned to a low value, +50 nN, in which case the surface morphology changes at a slow rate. Obviously, the slow process helps to confirm that the photoresist was continuously swept from right to left, in accordance with the retrace trajectory. For this reason, we always set the retrace trajectory as the stress writing direction.

**The stress writing process.** The commercial diamond coated probe (CDT-NCLR) was selected for the stress writing process due to the excellent wear resistant and a large force constant (~80 N/m). During the experiment, the probe works in the strong repulsive force status (+728 nN) and was driven to scan over the MoS$_2$ surface. Typically, the retrace-trajectory/writing-direction is fully controlled by adjusting the scanning area and angle. In this case, the tensile or compressive stress can be written into any sidewall of the triangular hole. For example, when the probe was controlled to climb up/down a certain sidewall on the retrace-journey, it will squeeze/drag the crystal lattice hard and then write compressive/tensile stress there.

**Spatially resolved differential-capacitance (SCM) and photocurrent (SPCM) experiments.** SCM is an approach developed based on AFM. In such measurement, the probe will not only record the surface morphology information, but also serves as a mobile nano-electrode to extract the differential-capacitance signal point by point. Considering that the sign and magnitude of differential-capacitance signal reflect the polarity and concentration of the doping carriers, SCM technique can measure the doping pattern of semiconductor in nanoscale resolution. To ensure the electrical contact of the measurements, a 15nm-Cr/45nm-Au electrode was deposited onto the MoS$_2$ surface ahead. During the measurements, this electrode was supplied with a 1V 90 kHz AC bias, while the probe was kept virtual ground.

SPCM images were obtained by applying a small laser beam to scan over the sample surface and recording the net photocurrent in real-time. Specifically, a 520 nm laser is focused onto the sample surface with ~2 μm in diameter and chopped at a frequency of 50 Hz. The laser beam is then controlled by a galvanometer (LSKGG4(/M), Thorlabs) to advance 200 nm at each step in both X and Y directions. The sample is wired to a current preamplifier (SR570) and a lock-in amplifier (SR830). The photocurrent is extracted at each step by locking the signal at the chopped frequency.

**Density Functional Calculations.** Density functional calculations were performed using the VASP code[41]. The electron-ion interaction was described by the frozen-core projector augmented wave (PAW) method[42]. The plane-wave cutoff for wave function expansion is 400 eV. The generalized gradient approximation of Perdew-Burke-Ernzerhof (GGA-PBE) was adopted as the exchange-correlation functional during the structural relaxation[43], and the convergence criterion is set to 0.02 eV/Å. The van der Waals interaction was included by using the correction scheme of Grimme[44]. To have a more accurate description on the electronic structure, the Heyd-Scuseria-Ernzerhof (HSE) hybrid functional was employed to calculate the defect charge transition level[45]. The defect charge-transition energy level $\epsilon(q/q')$ corresponds to the Fermi energy $E_F$ at which the formation energy for a defect α with different charge state q and q′ equals with each other. It can be calculated by:
$$\epsilon(q/q')=[E(\alpha,q)-E(\alpha,q')+(q-q')(E_{VBM}+\Delta V)]/(q'-q)^{46}$$
Here $E(\alpha,q)$ is the total energy of the supercell containing the defect, and $E_{VBM}$ is the valence band maximum (VBM) energy of the host material. $\Delta V$ is the correction for the potential alignment between the host and the supercell with a neutral defect. For charged defects, the Makov-Payne correction generalized to the anisotropic dielectric constant is also included to account for the spurious electrostatic interaction due to periodic boundary conditions[47]. A 6×6×2 supercell of $MoS_2$ containing 432 atoms is constructed for the defect calculations, and the Brillouin zone is sampled by the Gamma point.

**Sample fabrication for logic applications.** Standard lithography process (combined with etching process) was performed to depict a rectangular groove, with several micrometers in length, ~300 nm in width and ~70 nm in depth, on a Si/280-nm-$SiO_2$ substrate. A second-round lithography process was performed afterwards, followed by thermal evaporation of 45 nm Au film and lift-off process. The lithography window is strictly aligned to the groove region, which makes sure that the Au film deposits into the groove. After that, A uniform $MoS_2$ film is transferred onto the substrate and covers the etched area. The stress writing process was performed to squeeze the $MoS_2$ film into the groove and make it touch the bottom Au film. A third-round lithography and thermal evaporation processes were performed to deposit a 45nm Au film on the flat $MoS_2$ region (outside the groove). Those procedures finally result in a flexo-doped $MoS_2$ junction device. For the application of such device in logic OR gate, the flexo-doped $MoS_2$ junction is connected in parallel with a commercial silicon diode and loaded with a 1M resistance. The bottom and top contacts of the flexo-doped $MoS_2$ device are set as input and output ends, respectively.

**Ethics & Inclusion statement**

The research described in this manuscript doesn't include any animal experiments. The research described in this manuscript doesn't include any human research participants. We confirm that the research in this manuscript meets the Research Ethics outlined in this journal.